\documentclass[letterpaper, 10 pt, journal, twoside]{IEEEtran}
\let\authorrefmark\IEEEauthorrefmark

\usepackage{style}
\usepackage{comment}

\title{Robust Optimal Network Topology Switching for Zero Dynamics Attacks}

\author{Hiroyasu Tsukamoto\authorrefmark{1}\authorrefmark{3}, Joshua D. Ibrahim\authorrefmark{2}, Joudi Hajar\authorrefmark{2}, James Ragan\authorrefmark{2}, Soon-Jo Chung\authorrefmark{2}, Fred Y. Hadaegh\authorrefmark{2}\authorrefmark{3}
\thanks{\authorrefmark{1} Department of Aerospace Engineering, University of Illinois at Urbana-Champaign, Urbana, IL, {\tt\footnotesize \href{mailto:hiroyasu@illinois.edu}{hiroyasu@illinois.edu}}.}
\thanks{\authorrefmark{2} Division of Engineering and Applied Science, Caltech, Pasadena, CA, {\tt\footnotesize\{\href{mailto:jdibrahi@caltech.edu}{jdibrahi},
\href{mailto:jhajar@caltech.edu}{jhajar}, \href{mailto:jragan@caltech.edu}{jragan}, \href{mailto:sjchung@caltech.edu}{sjchung}, \href{mailto:hadaegh@caltech.edu}{hadaegh}\}@caltech.edu}.}
\thanks{\authorrefmark{3} Jet Propulsion Laboratory, Pasadena, CA {\tt\footnotesize
\{\href{mailto:hiroyasu.tsukamoto@jpl.nasa.gov}{hiroyasu.tsukamoto}, \href{mailto:fred.y.hadaegh@jpl.nasa.gov}{fred.y.hadaegh}\}@jpl.nasa.gov}.}
\thanks{This research  is funded by the Technology Innovation Institution (TII) under a contract with Caltech. The first author is funded by the University of Illinois at Urbana-Champaign. Part of the research was carried out at the Jet Propulsion Laboratory, California Institute of Technology, under a contract with the National Aeronautics and Space Administration.}}

\begin{document}
\maketitle

\begin{abstract}
The intrinsic, sampling, and enforced zero dynamics attacks (ZDAs) are among the most detrimental stealthy attacks in robotics, aerospace, and cyber-physical systems. They exploit internal dynamics, discretization, redundancy/asynchronous actuation and sensing, to construct disruptive attacks that are completely stealthy in the measurement. They work even when the systems are both controllable and observable. This paper presents a novel framework to robustly and optimally detect and mitigate ZDAs for networked linear control systems. We utilize controllability, observability, robustness, and sensitivity metrics written explicitly in terms of the system topology, thereby proposing a robust and optimal switching topology formulation for resilient ZDA detection and mitigation. Our main contribution is the reformulation of this problem into an equivalent rank-constrained optimization problem (i.e., optimization with a convex objective function subject to convex constraints and rank constraints), which can be solved using convex rank minimization approaches. The effectiveness of our method is demonstrated using networked double integrators subject to ZDAs.
\end{abstract}
\section{Introduction}
\label{sec_introduction}
Networked control systems find practical utility in diverse applications such as distributed optimization, power sharing in microgrids, clock synchronization in sensor networks, and robotic and aerospace multi-agent systems. This paper focuses on one of their significant security concerns, the zero dynamics attacks (ZDAs)~\cite{zda_discrete,zda_discrete2,secure_control,sampling_zda,enforced_zda,zda_tutorial}, which discreetly embed attack signals in the null space of control systems to alter the system behaviors maliciously. These attacks inherently cloak their impact by ensuring that the modified system output mirrors its unaltered counterpart, thereby posing great risks to the systems' stability, operation, and sustainability.
 
For linear systems, ZDAs can be classified mainly into three types: intrinsic ZDAs~\cite{zda_discrete,zda_discrete2,secure_control}, sampling ZDAs~\cite{sampling_zda}, and enforced ZDAs~\cite{enforced_zda,zda_tutorial}, which exploit specific characteristics of systems such as invariant zeros, discretization-induced unstable zeros, and system redundancies, respectively. All of these attacks can make the systems' states diverge completely stealthily, even if the systems are both controllable and observable. Their robust counterpart~\cite{robust_zda} further extends this idea to construct disruptive attacks for nonlinear systems and linear systems with uncertainty. It is also worth noting that the ZDAs have a tight connection to the zero dynamics and internal dynamics of~\cite{zero_dynamics_isidori}, which implies a great threat also in robotics and aerospace systems (\eg{}, hybrid robotics~\cite{hybrid_robotics}, PVTOL aircrafts~\cite{pvtol}, bipedal/quadrupedal robots~\cite{hybrid_robotics}, UAVs~\cite{zda_continuous_hovakimyan,uavzda}, and robotic manipulators~\cite{nmpzeros}).
\subsubsection*{Contributions}
Inheriting the performance and security metrics of existing literature (see, \eg{}, the controllability, observability, robustness, and sensitivity metrics of~\cite{gramian_original,10.1007/BFb0109870,attack_bullo,bullo2} and the references therein), we formulate our optimization problem for robust optimal detection and mitigation of the three types of ZDAs, explicitly expressed in terms of the networked linear control systems' \emph{communication topology}. Our main contribution is the reformulation of this problem into a mathematically equivalent rank-constrained optimization problem with the topology being a decision variable. This class of problems can be solved computationally efficiently with convex rank-minimization techniques~\cite{BoydRankRelaxation,rank_constrained_kyotoU,rank_constrained_boyd,rank_constrained_microsoft,rank_constrained_pourdue_ran}. We perform numerical simulations to demonstrate the effectiveness of our proposed robust optimal topology switching strategy for a network of double integrators. The results highlight how our method robustly and proactively averts ZDAs while keeping systems controllable and observable.
\subsubsection*{Related Work}
Numerous studies have focused on improving the reliability and security of networked control systems. \cite{security_metric_book,zda_discrete}~introduce security metrics that assess both the impact and detectability of attacks, informing the design of resilient observer-based controllers and detectors against stealthy attacks. \cite{attack_bullo,bullo2}~are another fundamental works that establish a theoretical foundation for robustly monitoring and detecting attacks in cyber-secure networked systems. As discussed extensively in these studies, a unique property of ZDAs is that they are not detectable through the systems' output by definition. This implies that the only counterstrategy is to change the dynamics somehow, \eg{}, using the generalized hold~\cite{sampling_zda}. Topology-switching strategies for ZDA detection are also based on this idea as proposed in~\cite{zda_discrete2}. \cite{zda_continuous_hovakimyan} further delves into defense tactics against novel ZDA-based stealthy attacks, where defenders switch between crafted network topologies to detect ZDAs white attackers simultaneously adapt to the dependers' strategy. As an example of practical applications, \cite{uavzda} considers ZDAs and covert attacks on networked UAVs in formation control, presenting decentralized and centralized detection schemes that leverage the communication topology switching of UAVs.
\subsubsection*{Notations}
\label{notation}
For $A \in \mathbb{R}^{n \times n}$, we use $A \succ 0$, $A \succeq 0$, $A \prec 0$, and $A \preceq 0$ for the positive definite, positive semi-definite, negative definite, negative semi-definite matrices, respectively. For $x \in \mathbb{R}^n$, we let $\|x\|$ denote the Euclidean norm.

\section{Preliminaries: Zero Dynamics Attacks and Problem Statement}
\label{sec_preliminaries}
We consider the following networked linear control system:
\begin{subequations}
\label{eq_system}
\begin{align}
    \dot{x}(t) &= A(t) x(t)+B(t,\nu(t))(u(t,y(t),\nu(t))+a(t))\label{eq_continuous_initialx}\\
    y(t) &= C(t,\nu(t))x(t)\label{eq_continuous_initialy}
\end{align}
\end{subequations}
where $t \in \mathbb{R}_{+}$, $\nu:\mathbb{R}_{+}\mapsto\mathcal{T}$ is the topology representation of the network (see Assumption~\ref{assump_affine_topology}) with $\mathcal{T}$ being the set of topologies, $x:\mathbb{R}_{+}\mapsto\mathbb{R}^p$ is the stacked system state, $u:\mathbb{R}_{+}\times\mathbb{R}^{r}\times\mathcal{T}\mapsto\mathbb{R}^q$ is the stacked system control input, $a:\mathbb{R}_{+}\mapsto\mathbb{R}^q$ is the stacked attack control input, $y:\mathbb{R}_{+}\mapsto\mathbb{R}^{r}$ is the stacked system measurement, and $A:\mathbb{R}_{+}\mapsto\mathbb{R}^{p\times p}$, $B:\mathbb{R}_{+}\times\mathcal{T}\mapsto\mathbb{R}^{p\times q}$, and $C:\mathbb{R}_{+}\times\mathcal{T}\mapsto\mathbb{R}^{{r}\times p}$ are the stacked system matrices.
\begin{remark}
\label{remark_estimation}
This formulation can be viewed as a general closed-loop dynamical system, where $x$ consists of the actual state, estimated state for feedback control, and estimated state for anomaly detection as in~\cite{outputL2gain}.
\end{remark}

In networked control systems in robotics, aerospace, and cyber-physical systems, switching the network topology $\nu$ does not change the open-loop system matrix $A$ of the system~\eqref{eq_continuous_initialx} as it is often determined by the underlying dynamics. The closed-loop $A$ can be modified by the feedback control $u$.
\begin{assumption}
\label{assump_system}
We assume the following in~\eqref{eq_system}.
\begin{enumerate}[label={\color{uiucblue}{\rm(\alph*)}}]
    \item The functions $A$, $B$, and $C$ are known both to the defender and the attacker.\label{item_ass1}
    \item The defender cannot directly measure the attack $a$.\label{item_ass2}
    \item The defender can switch the network topology for all $t$, and the attacker also knows the topology.\label{item_ass3}
    \item The actuation and sensing are performed discretely with the zero-order hold (see, \eg{},~\cite{zoh}), with $\Delta t_u$ and $\Delta t_y$ being the control actuation and sensing sampling period, respectively.\label{item_ass5}
    \item The network topology $\nu(t)$ is fixed during $t\in[k\Delta t_y,(k+1)\Delta t_y],~\forall k \in \mathbb{N}$.\label{item_ass6}
\end{enumerate}
\end{assumption}
\begin{assumption}[Network Topology]
\label{assump_affine_topology}
Let $\Theta_k \in \{0,1\}^{N\times N}$ denote the adjacency matrix of the network at each sensing time step $k$ of~\ref{item_ass6} in Assumption~\ref{assump_system}, where $N$ is the number of agents and $\{0,1\}^{N\times N}$ is the set of $N$ by $N$ matrices with their entries restricted to $0$ or $1$. The network topology $\nu$ of~\eqref{eq_system} is represented by $\Theta_k$, which describes the communication and physical connections of the network at each time step $k$. Also, the adjacency matrix $\Theta_k$ lies in a time-varying, finite set of matrices $\mathcal{T}_{k} \subset \mathcal{T}$ that represents all the possible distributed communications and physical connections of the network.
\end{assumption}
Assumption~\ref{assump_affine_topology} is mainly used for expressing the attack detection and attenuation problem in terms of the network topology $\nu$ in Sec.~\ref{sec_switching}. This is also useful for dealing with mobile agents with time-varying $\mathcal{T}_k$ due to distributed communications, as to be demonstrated in Sec.~\ref{sec_example}.
\subsection{Zero Dynamics Attack (ZDA)}
\label{sec_zda_definitions}
The ZDA is a stealthy attack designed based on the assumptions \ref{item_ass1} and \ref{item_ass2} of Assumption~\ref{assump_system}. In this paper, we define a stealthy (or undetectable~\cite{attack_bullo}) attack as follows.
\begin{definition}
\label{def_stealth}
For a given terminal time $t_F$, an attack $a(t)$ of~\eqref{eq_continuous_initialx} is stealthy in $t\in[0,t_F]$ if, for some initial states, we have $y(t) = y_n(t)$ for $\forall t\in[0,t_F]$, where $y$ is given by~\eqref{eq_continuous_initialy} and $y_n$ is the nominal unattacked output (\eqref{eq_continuous_initialy} with $a=0$).
\end{definition}
We focus on the ZDA as one of the most common stealthy attacks. It works for a wide range of real-world systems (\eg{}, robotics, aerospace, and cyber-physical systems) \textit{even when they are both controllable and observable}. Note that although our method deals with the ZDAs for linear time-varying systems, as will be seen in the following sections, some of the definitions below are given for linear time-invariant systems just to provide readers with high-level ideas of the ZDAs.
\subsubsection{Intrinsic ZDA}
This is a stealthy attack based on the invariant zeros $z_{\mathrm{inv}} \in \mathbb{C}$ of the system~\eqref{eq_system}, at which the following equation with the Rosenbrock system matrix
\begin{align}
    \label{eq_rosenbrock_cp}
    \begin{bmatrix}
    z_{\mathrm{inv}}\mathbb{I}-A & -B \\
    C & \mathbb{O}
\end{bmatrix}
\begin{bmatrix}
    x_{a0} \\ u_{a0}
\end{bmatrix} = 0
\end{align}
gives a nontrivial solution $[x_{a0}^{\top}~u_{a0}^{\top}]^{\top} \neq 0$, assuming that the system matrices are time-invariant and the network topology $\nu$ is fixed~\cite{zda_discrete,secure_control}. The intrinsic ZDA has a tight connection to the zero dynamics and internal dynamics of~\cite{zero_dynamics_isidori} that often appear in hybrid robotics~\cite{hybrid_robotics}. In fact, for linear systems, the system eigenvalues of the zero dynamics are equal to the invariant zeros of~\eqref{eq_rosenbrock_cp}~\cite[pp. 297-299]{Isidori:1995:NCS:545735}.
\begin{lemma}
\label{lemma_c_ZDA}
If there exist $z_{\mathrm{inv}}$, $x_{a0}$, $u_{a0}$ that satisfy~\eqref{eq_rosenbrock_cp} for a fixed network topology, an attack constructed as $a = e^{z_{\mathrm{inv}} t}u_{a0}$ with $x(0) = x_{n0}+x_{a0}$ is stealthy in the sense of Definition~\ref{def_stealth}, where $x_{n0}$ is a nominal initial state for~\eqref{eq_system} with $a=0$.
\end{lemma}
\begin{proof}
Let $X(s) = \mathcal{L}[x(t)]$, $U(s) = \mathcal{L}[u(t)]$, and $Y(s) = \mathcal{L}[y(t)]$ for the Laplace transform $\mathcal{L}$. Since we have $(s-z_{\mathrm{inv}})^{-1}(z_{\mathrm{inv}}\mathbb{I}-A) = -\mathbb{I}+(s-z_{\mathrm{inv}})^{-1}(s\mathbb{I}-A)$ for any $s$, applying $\mathcal{L}$ to~\eqref{eq_continuous_initialx} with the first row of~\eqref{eq_rosenbrock_cp} yields
\begin{align}
    X(s) = (s\mathbb{I}-A)^{-1}(x_{n0}+BU(s))+(s-z_{\mathrm{inv}})^{-1}x_{a0}.
\end{align}
The second row of~\eqref{eq_rosenbrock_cp} with~\eqref{eq_continuous_initialy} then gives $Y(s)= C(s\mathbb{I}-A)^{-1}(x_{n0}+BU(s))$ as desired.
\end{proof}

\begin{figure}
    \centering
    \includegraphics[height=24mm]{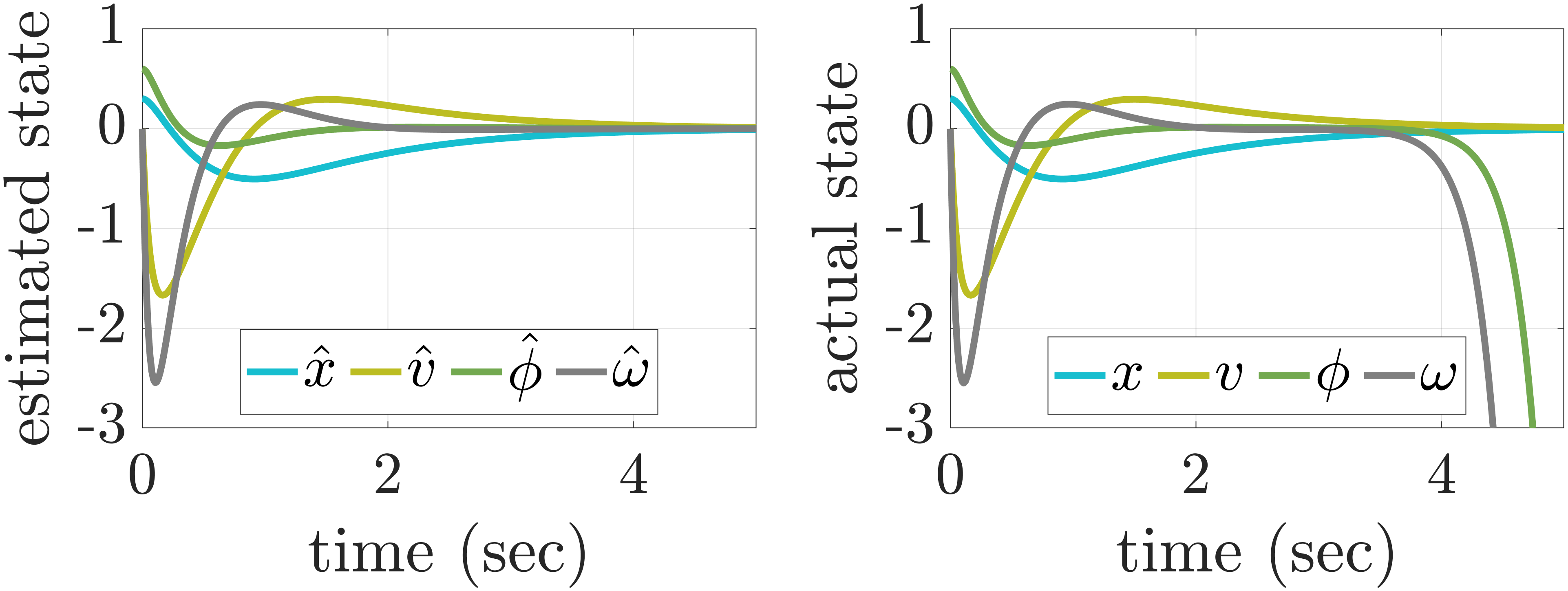} 
    \caption{Cart-pole balancing with the intrinsic ZDA, where $x$ is the cart's position, $v$ is the velocity, $\phi$ is the pole's angle from the upward equilibrium, and $\omega$ is the angular velocity as described \textcolor{uiucblue}{\href{https://ctms.engin.umich.edu/CTMS/index.php?example=InvertedPendulum&section=SystemModeling}{here}}. Due to the stealthy nature of the ZDA, the states estimated by the Kalman filter erroneously converge although the actual states diverge.}
    \label{fig_ex_cp_attack}\vspace{-20pt}
\end{figure}

Lemma~\ref{lemma_c_ZDA} indicates that the intrinsic ZDA stealthily makes $x$ of~\eqref{eq_continuous_initialx} diverge if the system is non-minimum-phase $\Leftrightarrow \operatorname{Re}(z_{\mathrm{inv}}) > 0 \Leftrightarrow$ the zero dynamics are unstable (see~\cite{zda_discrete,zda_discrete2} for discrete-time systems and \cite{attack_bullo,zda_continuous_hovakimyan} for continuous-time systems for more details). The zeros $z_{\mathrm{inv}}$ are ``invariant'' under coordinate changes and state feedback, which represent one of the ``intrinsic'' properties of zero dynamics and internal dynamics~\cite[pp. 218 -228]{Ref_Slotine}.
\begin{example}
\label{ex_cp}
The linearized system of the cart-pole balancing with the position measurement is subject to the ZDA, although it is both controllable and observable. This is part of the reason why the ZDA can be a great threat in robotics~\cite{hybrid_robotics} (see Fig.~\ref{fig_ex_cp_attack}). Some other examples include, but are not limited to, the PVTOL aircraft~\cite{pvtol}, bipedal/quadrupedal robots~\cite{hybrid_robotics}, UAVs~\cite{uavzda}, and robotic manipulators~\cite{nmpzeros}.
\end{example}
\subsubsection{Sampling ZDA}
Discretization of~\eqref{eq_system} could introduce additional potentially unstable zeros, called sampling zeros~\cite{zoh}. In fact, if the system's relative degree is $\geq 3$, we always have at least one unstable sampling zero~\cite[pp. 47-58]{yuz_sampled-data_2014} in addition to the intrinsic invariant zeros. The sampling ZDA, introduced in~\cite{sampling_zda}, essentially uses these zeros in the discretized system of~\eqref{eq_system}, which can be computed as in~\eqref{eq_rosenbrock_cp} by replacing its system matrices with the discretized counterparts. 
\begin{lemma}
\label{lemma_d_ZDA}
If $z_{\mathrm{inv}}$, $x_{a0}$, and $u_{a0}$ satisfy~\eqref{eq_rosenbrock_cp} for the discretized system and fixed network topology, an attack, $a = z_{\mathrm{inv}}^ku_{a0}$ with $x(0) = x_{n0}+x_{a0}$ is stealthy in the sense of Definition~\ref{def_stealth}, where $x_{n0}$ is a nominal initial state for~\eqref{eq_system} with $a=0$.
\end{lemma}
\begin{proof}
This follows from taking the Z-transform instead of the Laplace transform for the discretized system in Lemma~\ref{lemma_c_ZDA}.
\end{proof}
Lemma~\ref{lemma_d_ZDA} indicates that the sampling ZDA stealthily makes $x$ of~\eqref{eq_continuous_initialx} diverge if the discretized system is non-minimum-phase $\Leftrightarrow |z_{\mathrm{inv}}| > 1 \Leftrightarrow$ the discretized zero dynamics is unstable. Some examples of sampling zeros are given in~\cite{zda_tutorial} and the references therein.
\subsubsection{Enforced ZDA}
Even if the discretized system is minimum-phase (\ie, there are no unstable zeros), the attacker can still easily design a disruptive stealthy attack by exploiting actuation redundancy or the asynchronous actuation and sensing of~\ref{item_ass5} in Assumption~\ref{assump_system}. Such an attack is called the enforced ZDA~\cite{enforced_zda,zda_tutorial}. This significantly broadens the applicability of ZDAs to controllable, observable, and minimum-phase real-world systems with discrete-time actuation and sensing.
\subsection{Motivation and Problem Statement}
\begin{figure}
    \centering
    \includegraphics[height=24mm]{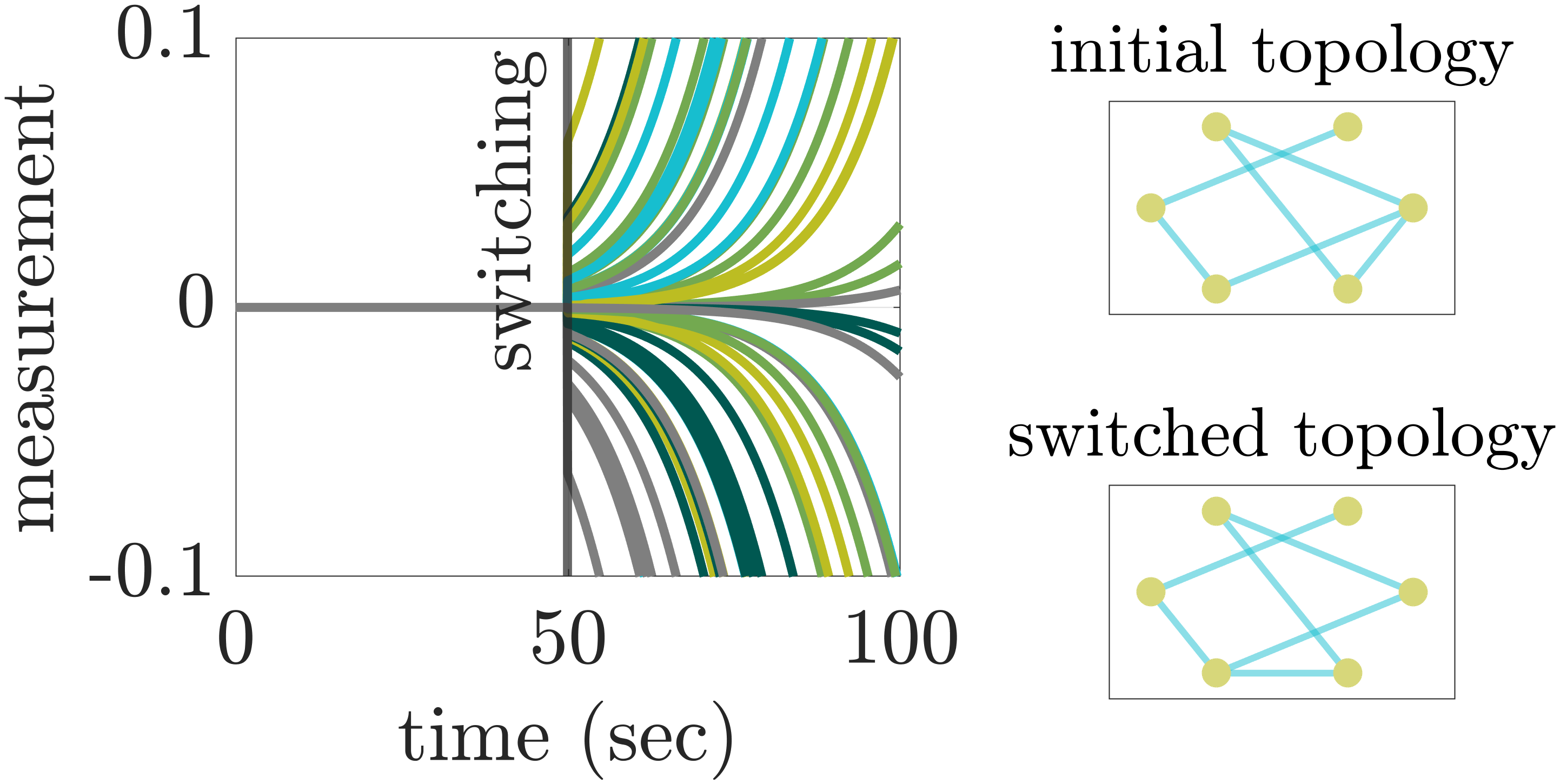} 
    \caption{Networks of 3D double integrators with the enforced ZDA. Although the attack is stealthy initially, the network topology switching reveals it at $t = 50$ (sec).}
    \label{fig_ex_dinet_attack}\vspace{-20pt}
\end{figure}

As implied in Sec.~\ref{sec_zda_definitions}, detecting and mitigating ZDAs require changing the underlying dynamics of~\eqref{eq_system}, which leads to our motivation for robust and optimal topology switching.
\begin{example}
\label{ex_dinet}
Consider the network of six 3D double integrators with the enforced ZDA to see the effectiveness of network topology switching. We consider the two topologies given on the right-hand side of Fig.~\ref{fig_ex_dinet_attack}, both of which are controllable, observable, and minimum-phase with the full position measurements, with a position feedback synchronization controller. The enforced ZDA is indeed completely stealthy initially, but switching the network topology makes the attack visible in the measurement (see Fig.~\ref{fig_ex_dinet_attack})
\end{example}

Switching the network topology indeed helps us detect ZDAs, as implied in Example~\ref{ex_dinet}. Let us formalize this observation to further motivate this idea, taking the intrinsic ZDA of~\eqref{eq_rosenbrock_cp} as an example.
\begin{lemma}
\label{lemma_output_eig}
The input $u_{a 0} = F x_{a 0}$ and $x_{a 0} \in \mathcal{V}^{\star}$ such that $(A+B F) \mathcal{V}^{\star} \subseteq \mathcal{V}^{\star} \subseteq \operatorname{ker}(C)$ are the eigenvalues of $A+B F$ restricted to the eigenspace spanned by $\mathcal{V}^{\star}$ and correspond to the invariant zeros of \eqref{eq_rosenbrock_cp} yielding $y=0, \forall t \geq 0$. 
\end{lemma}
\begin{proof}
Let $\lambda\left(\left.M\right|_{\mathcal{V}}\right)$ denote the eigenvalues of some matrix $M$ whose eigenvectors belong to the subspace $\mathcal{V}$. 
Substituting $u_{a 0}=F x_{a 0}$ into~\eqref{eq_rosenbrock_cp}, we can see that \eqref{eq_rosenbrock_cp} with $u_{a 0}=F x_{a 0}$ is satisfied when the tuple $\left(z_{\text {inv }}, u_{a 0}, x_{a 0}\right)$ is chosen such that $z_{\text {inv }} \in \lambda\left(\left.(A+B F)\right|_{\mathcal{V}^{\star}}\right)$ and $x_{a 0} \in \mathcal{V}^{\star}$. See~\cite{zda_discrete2} for details.
\end{proof}

The geometric interpretation of the intrinsic ZDA in Lemma~\ref{lemma_output_eig}, originally introduced in~\cite{zda_discrete2}, implies that 1) when a new topology $\tilde v$ induces a change in matrix $C$ to $\tilde C$, a intrinsic ZDAs are revealed if only if $\mathcal{V}^{\star} \cap \operatorname{ker}(\tilde{C}) = \emptyset$, and 2) when $\tilde v$ induces a change in $B$ to $\tilde B$, intrinsic ZDAs are revealed if and only if $\mathcal{V}^{\star} \cap \operatorname{ker}((\tilde{B}-B) F)=\emptyset$.

Our question is, how can we optimally perform such network topology switching so the defender can robustly detect and mitigate the disruptive ZDAs of Sec.~\ref{sec_zda_definitions}? Note that there exists a robust version of ZDAs in literature~\cite{robust_zda}, which implies an extension of our approach to nonlinear systems and linear systems with uncertainties.

\section{Preliminaries: Performance Metrics for Switching}
\label{sec_metrics}
In order to quantify the quality of the network topology switching as in Example~\ref{ex_dinet}, this section defines several performance metrics expressed in terms of the network topology $\nu$ of~\eqref{eq_system}. Most of the results here are preliminaries but explicitly stated here just to make this paper as self-contained as possible. Those familiar with the topics to be covered here could skip this section and proceed to our main contribution in Section~\ref{sec_switching}.

We consider finite-horizon metrics so we can handle time-varying ZDA detection and mitigation strategies.
\begin{remark}
\label{ref_remnote}
    In the subsequent discussion, we suppose that Assumption~\ref{assump_system} holds unless stated otherwise. Also, we let $\ell$ and $k$ be the sampling indices for the control and sensing, respectively, $t_F$ be the terminal time, $x_S$ is the initial state, $x_F$ is the terminal state, $t_{\ell} = \ell\Delta t_u$ and $t_{k} = k\Delta t_y$ for the time periods of~\ref{item_ass5}, $\Ell = \floor{t_F/\Delta t_u}$ and $K = \floor{t_F/\Delta t_y}$ for the floor function $\floor{\cdot}$, and $u_{\ell} = u(t_{\ell},y(t_\ell),\nu(t_{\ell}))$, $a_{\ell} = a(t_{\ell})$, $x_k = x(t_k)$, $A_k = A(t_k)$, $B_{k} = B(t_k,\nu(t_k))$, and $C_k = C(t_k,\nu(t_k))$ in the system~\eqref{eq_system}.
\end{remark}
\subsection{Discretization with Asynchronous Actuation and Sensing}
Before we proceed, let us compute the discretized counterparts of~\eqref{eq_system} with the asynchronous actuation and sensing of Assumption~\ref{assump_system}. This is achieved simply by using the impulse response $h_{r}(t)$ of the zero-order hold, given as $h_{r}(t) = 1$ if $t \in [0,\Delta t_{u})$ \& $h_{r}(t) = 0$ otherwise, which results in the control and attack input, $\mu\in\{u,a\}$, expressed as 
\begin{align}
    \label{eq_mu_def}
    \mu(t) = \sum_{\ell=0}^{\Ell}h_{r}(t-t_\ell)\mu_\ell.
\end{align}
\begin{lemma}
\label{lemma_discretization}
We define $\mathbf{B}_{k,\Ell}$ as $= [H_{k,0}B_{k},\cdots,H_{k,\Ell}B_{k}]$ using the notation in Remark~\ref{ref_remnote}, where $H_{k,\ell} = \int_{t_k}^{t_{k+1}}e^{A_k(t_{k+1}-\tau)}h_{r}(\tau-t_\ell)d\tau$. If \ref{item_ass5}~and~\ref{item_ass6} of Assumption~\ref{assump_system} hold, \eqref{eq_continuous_initialx} is discretized as:
\begin{align}
    \label{eq_all_state}
    \mathbf{x}_{K} &= \mathcal{A}_{K}x_S+\mathcal{B}_{K,\Ell}(\mathbf{u}_{\Ell}+\mathbf{a}_{\Ell})
\end{align}
where $S_k = e^{A_k\Delta t_y}$, $\mathcal{A}_{K} = [\mathbb{I}^{\top},S_0^{\top},\cdots,(S_{K-1}\cdots S_0)^{\top}]^{\top}$, $\mathbf{x}_K = [x_0^{\top},\cdots,x_K^{\top}]^{\top}$, $\mathbf{u}_{\Ell} = [u_0^{\top},\cdots,u_{\Ell}^{\top}]^{\top}$, $\mathbf{a}_{\Ell} = [a_{0}^{\top},\cdots,a_{\Ell}^{\top}]^{\top}$, and
\begin{align}
    \mathcal{B}_{K,\Ell} = 
    \left[\begin{smallmatrix}
        \mathbb{O} \\
        \mathbf{B}_{0,L} \\
        S_1\mathbf{B}_{0,L}+\mathbf{B}_{1,L} \\
        \vdots \\
        S_{K-1}\cdots S_1\mathbf{B}_{0,L}+\cdots+\mathbf{B}_{K-1,L}
    \end{smallmatrix}\right].~\label{eq_Bmathcaldef}
\end{align}
Also, using the block diagonal matrix $\blkdiag{(\cdot)}$, the measurement~\eqref{eq_continuous_initialy} can be computed as follows:
\begin{align}
    \label{eq_all_measurement}
    \mathbf{y}_K = \mathcal{C}_K\mathcal{A}_{K}x_S+\mathcal{C}_K\mathcal{B}_{K,\Ell}(\mathbf{u}_{\Ell}+\mathbf{a}_{\Ell})
\end{align}
where $\mathcal{C}_K = \blkdiag{(C_0,\cdots,C_K)}$ and $\mathbf{y}_K = [y_0^{\top},\cdots,y_K^{\top}]^{\top}$.
\end{lemma}
\begin{proof}
Integrating~\eqref{eq_system} with $u(t)$ and $a(t)$ given by~\eqref{eq_mu_def} completes the proof.
\end{proof}
\subsection{Transient Controllability and Observability Metrics}
The minimum control effort to reach the terminal state $x_F$ from the initial state $x_S$, which characterizes controllability, can be computed as follows assuming $\mathbf{a}_{\Ell} = 0$ in~\eqref{eq_all_state}~\cite{gramian_original}:
\begin{align}
    \label{eq_controllability_obj}
    \bar{J}_{\mathrm{con}} = \inf_{\mathbf{u}_{\Ell}}{\mathbf{u}_{\Ell}^{\top}\mathbf{u}_{\Ell}}~\mathrm{\st}~\mathbf{x}_{K} = \mathcal{A}_{K}x_S+\mathcal{B}_{K,\Ell}\mathbf{u}_{\Ell},~x_K=x_F 
\end{align}
where the notation is consistent with~\eqref{eq_all_state}.
\begin{lemma}
\label{lemma_controllability}
If $(\mathcal{B}_{K,\Ell})_K(\mathcal{B}_{K,\Ell})_K^{\top}$ is invertible, the metric of~\eqref{eq_controllability_obj} is given as $\bar{J}_{\mathrm{con}} = \Delta x_{SF}^{\top}\left((\mathcal{B}_{K,\Ell})_K(\mathcal{B}_{K,\Ell})_K^{\top}\right)^{-1}\Delta x_{SF}$, where $\Delta x_{SF} = x_F-(\mathcal{A}_K)_Kx_S$, $(\mathcal{A}_K)_K$ and $(\mathcal{B}_{K,\Ell})_K$ are the last $n$ rows of $\mathcal{A}_K$ and $\mathcal{B}_{K,\Ell}$, respectively, and the other notations are given in~\eqref{eq_all_state} and in Remark~\ref{ref_remnote}.
\end{lemma}
\begin{proof}
Since the problem~\eqref{eq_controllability_obj} is convex, applying the KKT condition completes the proof.
\end{proof}

We thus define the following transient controllability metric as a result of Lemma~\ref{lemma_controllability}:
\begin{align}
    \label{eq_controllability_explicit_final}
    J_{\mathrm{con}}(\nu) = \lambda_{\min}\left((\mathcal{B}_{K,\Ell})_K(\mathcal{B}_{K,\Ell})_K^{\top}\right)
\end{align}
where $\lambda_{\min}(\cdot)$ denotes the minimum eigenvalue.
\begin{remark}
When the actuation and sensing frequencies are equal, the term $(\mathcal{B}_{K,\Ell})_K(\mathcal{B}_{K,\Ell})_K^{\top}$ in~\eqref{eq_controllability_explicit_final} reduces to the controllability Gramian~\cite{gramian_original} and can be computed recursively.
\end{remark}

Similarly, assuming $\mathbf{u}_{\Ell} = \mathbf{a}_{\Ell} = 0$ in~\eqref{eq_all_measurement}, the observability metric~\cite{gramian_original} is characterized by $\bar{J}_{\mathrm{obs}} = \mathbf{y}_K^{\top}\mathbf{y}_K~\mathrm{\st}~\mathbf{y}_K = \mathcal{C}_K\mathcal{A}_{K}x_S$, which gives $\bar{J}_{\mathrm{obs}} = x_S^{\top}\mathcal{A}_{K}^{\top}\mathcal{C}_K^{\top}\mathcal{C}_K\mathcal{A}_{K}x_S$, where the notation is consistent with~\eqref{eq_all_measurement}. Since observability is not affected by the asynchronous nature discussed in Lemma~\ref{lemma_discretization}, this corresponds to the observability Gramian as expected. We thus define the transient observability metric as
\begin{align}
    \label{eq_observability_explicit_final}
    J_{\mathrm{obs}}(\nu) = \lambda_{\min}\left(\mathcal{A}_{K}^{\top}\mathcal{C}_K^{\top}\mathcal{C}_K\mathcal{A}_{K}\right)
\end{align}
where $\lambda_{\min}(\cdot)$ denotes the minimum eigenvalue.
\subsection{Transient Attack Robustness Metric}
This metric can be defined with the following transient $\mathcal{H}_{\infty}$ norm (or $\mathcal{L}_2$ gain in a nonlinear sense)~\cite[Sec.~5.3]{Khalil:1173048} from the attack $\mathbf{a}_{\Ell}$ and the initial state $x_S$ to the full state $\mathbf{x}_K$, assuming $\mathbf{u}_{\Ell} = 0$ in~\eqref{eq_all_state}:
\begin{subequations}
\label{eq_robustness}
\begin{align}
    &J_{\mathrm{rob}}(\nu) = \sup_{x_S,\mathbf{a}_{\Ell}}\mathbf{x}_K^{\top}\mathbf{Q}_{\mathrm{rob}}\mathbf{x}_K \label{eq_robustness_obj} \\
    &\mathrm{\st}~\mathbf{x}_{K} = \mathcal{A}_{K}x_S+\mathcal{B}_{K,\Ell}\mathbf{a}_{\Ell},~\|x_S\|^2+\|\mathbf{a}_{\Ell}\|^2 \leq 1 \label{eq_robustness_con}
\end{align}
\end{subequations}
where the notation is consistent with~\eqref{eq_all_state}, $\mathbf{Q}_{\mathrm{rob}}\succ0$ is the weight on the state, and $\|x_S\|^2+\|\mathbf{a}_{\Ell}\|^2 \leq 1$ is introduced here to avoid the trivial solution of $\mathbf{a}_{\Ell} = \infty$ as in the operator norm.
\begin{lemma}
\label{lemma_robustness}
The attack robustness metric of~\eqref{eq_robustness} is given as
\begin{align}
    \label{eq_robust_explicit}
    J_{\mathrm{rob}}(\nu) = \lambda_{\max}\left(
    \left[\begin{smallmatrix}
        \mathcal{A}_{K}^{\top} \\ \mathcal{B}_{K,\Ell}^{\top}
    \end{smallmatrix}\right]
    \mathbf{Q}_{\mathrm{rob}}
    \left[\begin{smallmatrix}
        \mathcal{A}_{K} & \mathcal{B}_{K,\Ell}
    \end{smallmatrix}\right]\right)
\end{align}
where $\lambda_{\max}(\cdot)$ denotes the maximum eigenvalue and the other notations are given in~\eqref{eq_all_state} and in Remark~\ref{ref_remnote}.
\end{lemma}
\begin{proof}
Using the first constraint of~\eqref{eq_robustness_con}, we get $\mathbf{x}_K^{\top}\mathbf{Q}_{\mathrm{rob}}\mathbf{x}_K = [x_S^{\top},\mathbf{a}_{\Ell}^{\top}][\mathcal{A}_{K},\mathcal{B}_{K,\Ell}]^{\top}\mathbf{Q}_{\mathrm{rob}}[\mathcal{A}_{K},\mathcal{B}_{K,\Ell}][x_S^{\top},\mathbf{a}_{\Ell}^{\top}]^\top$. Since we have $[\mathcal{A}_{K},\mathcal{B}_{K,\Ell}]^{\top}\mathbf{Q}_{\mathrm{rob}}[\mathcal{A}_{K},\mathcal{B}_{K,\Ell}]\succeq 0$, upper-bounding this relation using the second constraint of~\eqref{eq_robustness_con} yields~\eqref{eq_robust_explicit}. The fact that the bound is the supremum in~\eqref{eq_robustness_obj} can be shown by contradiction~\cite[p. 210]{Khalil:1173048}
\end{proof}
\begin{remark}
Although we consider the controllability and attack robustness separately to access each property individually, we could also combine them to get the game-theoretic controllability Gramian~\cite{Basar1995,learning_actuator_selection}.
\end{remark}
\subsection{Transient Minimum Attack Sensitivity Metric}
This is the most important metric for the detection of the ZDAs introduced in~\eqref{sec_zda_definitions}, which measures the stealthiness of malicious attacks in the sense of Definition~\ref{def_stealth}. Given an initial state $x_S$, this metric can be defined with the following transient $\mathcal{H}_{-}$ index~\cite{security_metric_book,Hsubobserver,LMIsensitivity,secure_control,outputL2gain}, assuming $\mathbf{u}_{\Ell} = 0$ in~\eqref{eq_all_measurement}:
\begin{subequations}
\label{eq_sensitivity}
\begin{align}
    &J_{\mathrm{sen}}(\nu) = \inf_{x_S,\mathbf{a}_{\Ell}}\mathbf{y}_K^{\top}\mathbf{y}_K \label{eq_sensitivity_obj} \\
    &\mathrm{\st}~
    \mathbf{y}_K = \mathcal{C}_K\mathcal{A}_{K}x_S+\mathcal{C}_K\mathcal{B}_{K,\Ell}\mathbf{a}_{\Ell},~\|x_S\|^2+\|\mathbf{a}_{\Ell}\|^2 \geq 1\label{eq_sensitivity_con}
\end{align}
\end{subequations}
where the notation is consistent with~\eqref{eq_all_measurement} and $\|x_S\|^2+\|\mathbf{a}_{\Ell}\|^2 \geq 1$ is introduced here to avoid the trivial solution of $x_S = 0$ and $\mathbf{a}_{\Ell} = 0$ in the ZDA~\eqref{eq_rosenbrock_cp}.
\begin{lemma}
\label{lemma_sensitivity}
The minimum attack sensitivity metric of~\eqref{eq_sensitivity} is 
\begin{align}
    \label{eq_sensitivity_explicit}
    J_{\mathrm{sen}}(\nu) = \lambda_{\min}\left(
    \left[\begin{smallmatrix}
        \mathcal{A}_{K}^{\top} \\ \mathcal{B}_{K,\Ell}^{\top}
    \end{smallmatrix}\right]
    \mathcal{C}_K^{\top}
    \mathcal{C}_K
    \left[\begin{smallmatrix}
        \mathcal{A}_{K} & \mathcal{B}_{K,\Ell}
    \end{smallmatrix}\right]\right)
\end{align}
where $\lambda_{\min}(\cdot)$ denotes the minimum eigenvalue and the other notations are given in~\eqref{eq_all_measurement} and in Remark~\ref{ref_remnote}.
\end{lemma}
\begin{proof}
\eqref{eq_sensitivity_explicit} follows from lower-bounding $\mathbf{y}_K^{\top}\mathbf{y}_K$ using~\eqref{eq_sensitivity_con} as in the proof of Lemma~\ref{lemma_robustness}. The bound can be shown to be the infimum in~\eqref{eq_sensitivity_obj} by contradiction~\cite[p. 210]{Khalil:1173048}.
\end{proof}
\section{Main Contribution: Robust and Optimal Network Topology Switching}
\label{sec_switching}
Now that we have metrics for quantifying the key system performances, this section presents several numerically efficient optimization approaches for designing switching strategies of the network topology. We still suppose that Assumption~\ref{assump_system} holds, and we use the notation of Remark~\ref{ref_remnote}.
\subsection{Controlling Network Topology}
Feedbacking the network output as follows provides additional control authority over the modification of the system matrices:
\begin{align}
    \label{eq_feedback_control}
    \mathbf{u}_{\Ell} &=  \mathcal{K}_{\Ell,K}\mathbf{y}_K+\mathbf{v}_{\Ell}= 
    \left[\begin{smallmatrix}
    f_{s}(t_0-t_0)\kappa_{0,0} & \cdots & f_{s}(t_0-t_K)\kappa_{0,K} \\
    \vdots & \ddots & \vdots \\
    f_{s}(t_{\Ell}-t_0)\kappa_{\Ell,0} & \cdots & f_{s}(t_{\Ell}-t_K)\kappa_{\Ell,K}
    \end{smallmatrix}\right]\mathbf{y}_K+\mathbf{v}_{\Ell}
\end{align}
where $\mathbf{y}_K$ and $\mathbf{u}_{\Ell}$ are as given in Lemma~\ref{lemma_discretization}, $\mathbf{v}_{\Ell}$ is a residual control input for additional tasks, $\kappa_{\ell,k} \in \mathbb{R}^{q\times p}$ are the control gains, and $f_{s}$ is the unit step function, \ie{}, $f_{s}(t) = 1$ if $t\geq 0$ and $f_{s}(t) = 0$ otherwise. The step functions are just to ensure the causality of the state feedback.
\begin{remark}
The output feedback of the form~\eqref{eq_feedback_control} could also include systems with state estimation by introducing internal states as in the Kalman filter. See~\cite{sls_output_feedback_control,outputL2gain} for more details.
\end{remark}

Using the feedback control policy~\eqref{eq_feedback_control} in~\eqref{eq_all_state}, we have
\begin{align}
    \label{eq_control_substitute}
    (\mathbb{I}-\mathcal{B}_{K,\Ell}\mathcal{K}_{\Ell,K}\mathcal{C}_{K})\mathbf{x}_{K} &= \mathcal{A}_{K}x_S+\mathcal{B}_{K,\Ell}(\mathbf{v}_{\Ell}+\mathbf{a}_{\Ell}).
\end{align}
Lemma~\ref{lemma_Kinvertible} explicitly derives $\mathbf{x}_{K}$ under feedback control.
\begin{lemma}
\label{lemma_Kinvertible}
The matrix $\mathcal{L}_{\Ell,K} = \mathbb{I}-\mathcal{B}_{K,\Ell}\mathcal{K}_{\Ell,K}\mathcal{C}_{K}$ of~\eqref{eq_control_substitute} is invertible and thus we have the following:
\begin{align}
    \label{eq_control_state_explicit}
    \mathbf{x}_{K} &= \mathcal{L}_{\Ell,K}^{-1}(\mathcal{A}_{K}x_S+\mathcal{B}_{K,\Ell}(\mathbf{v}_{\Ell}+\mathbf{a}_{\Ell})).
\end{align}
\end{lemma}
\begin{proof}
By definition of $H_{k,\ell}$ in Lemma~\ref{lemma_discretization} and the step function $f_{s}$ in~\eqref{eq_feedback_control}, we have $H_{k,\ell}f_{s}(t_{\ell}-t_{j})K_{\ell,j} = 0~\mathrm{if}~k+1 \leq j.$ This indicates that the matrix $\mathcal{B}_{K,\Ell}\mathcal{K}_{\Ell,K}\mathcal{C}_{K}$ is strictly lower triangular for $\mathcal{B}_{K,\Ell}$ of~\eqref{eq_Bmathcaldef}, $\mathcal{C}_{K}$ of~\eqref{eq_all_measurement}, and $\mathcal{K}_{\Ell,K}$ of~\eqref{eq_feedback_control}, which can be shown also by the system's causality. The matrix $\mathcal{L}_{\Ell,K}$ is thus invertible and~\eqref{eq_control_state_explicit} then follows from~\eqref{eq_control_substitute}.
\end{proof}

Lemma~\ref{lemma_Kinvertible} also leads to the following, which decouples the feedback control from the minimum-attack sensitivity of~\eqref{eq_sensitivity_explicit}.
\begin{lemma}
\label{lemma_feedback_decoupling}
The stealthiness of the system~\eqref{eq_system} in the sense of Definition~\ref{def_stealth} is invariant under the feedback control policy~\eqref{eq_feedback_control} with $\mathbf{v}_{\Ell} = 0$, \ie{}, if $\exists \mathcal{K}_{\Ell,K}$ \st{} $J_{\mathrm{sen}}(\nu) > 0$, then $J_{\mathrm{sen}}(\nu) > 0,~\forall \mathcal{K}_{\Ell,K}$, and if $\exists \mathcal{K}_{\Ell,K}$ \st{} $J_{\mathrm{sen}}(\nu) = 0$, then $J_{\mathrm{sen}}(\nu) = 0,~\forall \mathcal{K}_{\Ell,K}$, where $J_{\mathrm{sen}}(\nu)$ is the minimum attack sensitivity metric of Lemma~\ref{lemma_sensitivity}.
\end{lemma}
\begin{proof}
Substituting $\mathbf{u}_{\Ell} = \mathcal{K}_{\Ell,K}\mathbf{y}_{K}$ into~\eqref{eq_all_state} and then into~\eqref{eq_all_measurement} gives $\mathbf{y}_K = \mathcal{C}_{K}\mathcal{L}_{\Ell,K}^{-1}[\mathcal{A}_{K},\mathcal{B}_{K,\Ell}][x_S^{\top},\mathbf{a}_{\Ell}^{\top}]^{\top}$, as a result of Lemma~\ref{lemma_Kinvertible}.
The existence of the stealthy attacks given in Definition~\ref{def_stealth} is thus determined by the existence of the null space of $\mathcal{C}_{K}\mathcal{L}_{\Ell,K}^{-1}[\mathcal{A}_{K},\mathcal{B}_{K,\Ell}]$ as in Lemma~\ref{lemma_sensitivity}. Since the feedback part $\mathcal{L}_{\Ell,K} = \mathbb{I}-\mathcal{B}_{K,\Ell}\mathcal{K}_{\Ell,K}\mathcal{C}_{K}$ is nonsingular, the existence of the null space is invariant under feedback, which is consistent with the fact that the invariant zeros of the Rosenbrock matrix~\eqref{eq_rosenbrock_cp} is invariant under feedback.
\end{proof}
\subsection{Convex Formulation}
\label{sec_convex_formulation}
Controllability, observability, and non-stealthiness (given in Definition~\ref{def_stealth}) can be guaranteed as long as the controllability, observability, and minimum attack sensitivity metrics, respectively, are positive, and thus they are best used as constraints. Also, controllability, observability, and non-stealthiness are invariant under feedback, the third of which is shown in Lemma~\ref{lemma_feedback_decoupling}. The optimization problem for minimizing the robustness metric of Lemma~\ref{lemma_robustness} under these constraints can be formulated as follows:
\begin{align}
    \label{eq_switching_nonlinear}
    &\min_{\nu\in\mathcal{T}}~J_{\mathrm{rob}}(\nu)\mathrm{~\st{}~}J_{\mathrm{con}}(\nu) \geq c_{c},~J_{\mathrm{obs}}(\nu) \geq c_{o},\mathrm{~and~}J_{\mathrm{sen}}(\nu) \geq c_{s}
\end{align}
where $\nu$ is the network topology, $J_{\mathrm{con}}$, $J_{\mathrm{obs}}$, $J_{\mathrm{rob}}$, and $J_{\mathrm{sen}}$ are defined in Lemmas~\ref{lemma_controllability}~--~\ref{lemma_sensitivity}, $c_{c}$, $c_{o}$, and $c_{s}$ are some user-defined constants, and $J_{\mathrm{rob}}(\nu)$ is computed for the closed-loop system~\eqref{eq_control_state_explicit}, \ie{},
\begin{align}
    J_{\mathrm{rob}}(\nu) = \lambda_{\max}\left(
    \left[\begin{smallmatrix}
        \mathcal{A}_{K}^{\top} \\ \mathcal{B}_{K,\Ell}^{\top}
    \end{smallmatrix}\right]
    (\mathcal{L}_{\Ell,K}^{-1})^{\top}\mathbf{Q}_{\mathrm{rob}}(\mathcal{L}_{\Ell,K}^{-1})
    \left[\begin{smallmatrix}
        \mathcal{A}_{K} & \mathcal{B}_{K,\Ell}
    \end{smallmatrix}\right]\right).\nonumber
\end{align}

We now present the main result of this paper, which shows that the problem~\eqref{eq_switching_nonlinear} can be written as rank-constrained optimization with a linear objective and with linear matrix inequalities (LMIs) in terms of the system matrices and switching control gain.
\begin{theorem}
\label{thm_Afixed}
Suppose that the system matrix $A$ of~\eqref{eq_continuous_initialx} is constant at each time step $k$. If we view the system matrices of Lemma~\ref{lemma_discretization} and the control gain $\mathcal{K}_{\Ell,K}$ of Lemma~\ref{lemma_Kinvertible} as decision variables, the problem~\eqref{eq_switching_nonlinear} can be reformulated to the following equivalent optimization problem, \ie{}, $\gamma^* = \min J_{\mathrm{rob}}$ for $J_{\mathrm{rob}}$ of~\eqref{eq_switching_nonlinear}, where the objective function and constraints are all convex except for the rank constraints~\eqref{eq_switching_almost_bmi_con5}:
\begin{subequations}
\label{eq_switching_almost_bmi}
\begin{align}
    \label{eq_switching_almost_bmi_objective}
    &\gamma^* = \min_{\substack{\mathcal{X}_{b},\mathcal{X}_{c},\mathcal{X}_{k},\mathcal{X}_{bc},\mathcal{X}_{bk}\\\mathcal{B}_{K,\Ell},\mathcal{C}_{K},\mathcal{K}_{\Ell,K}}}~\gamma\\
    &\mathrm{~\st{}~}\phi_{b}(\mathcal{X}_{b}) \succeq c_{c}\mathbb{I},~\mathcal{A}_{K}^{\top}\Phi_{c}(\mathcal{X}_{c})\mathcal{A}_{K} \succeq c_{o}\mathbb{I}, \label{eq_switching_almost_bmi_con1}\textcolor{white}{\eqref{eq_switching_almost_bmi_con1}} \\
    &\textcolor{white}{\mathrm{\st{}~}}\left[\begin{smallmatrix}
    \mathcal{A}_{K}^{\top}\Phi_{c}(\mathcal{X}_{c})\mathcal{A}_{K} & \mathcal{A}_{K}^{\top}\Phi_{c}(\mathcal{X}_{c})\mathcal{B}_{K,\Ell}  \\
    \mathcal{B}_{K,\Ell}^{\top}\Phi_{c}(\mathcal{X}_{c})\mathcal{A}_{K} & \Psi_{bc}(\mathcal{X}_{bc})
    \end{smallmatrix}\right] \succeq c_{s}\mathbb{I}, \label{eq_switching_almost_bmi_con2}\textcolor{white}{\eqref{eq_switching_almost_bmi_con2}} \\
    &\textcolor{white}{\mathrm{\st{}~}}
    \left[\begin{smallmatrix}
    \mathcal{Y}_{bk} & \mathcal{E}_{K,\Ell} & \mathcal{F}_{K,\Ell} \\
    \mathcal{E}_{K,\Ell}^{\top} & \gamma\mathbb{I} & \mathbb{O} \\
    \mathcal{F}_{K,\Ell}^{\top} &\mathbb{O} & \mathbb{I}
    \end{smallmatrix}\right] \succeq 0, \label{eq_switching_almost_bmi_con3}\textcolor{white}{\eqref{eq_switching_almost_bmi_con3}} \\
    &\textcolor{white}{\mathrm{\st{}~}}
    \left[\begin{smallmatrix}
        \mathcal{X}_{\xi} & \matvec(\Xi) \\
        \matvec(\Xi)^{\top} &1
    \end{smallmatrix}\right] \succeq 0,~\forall(\xi,\Xi) \label{eq_switching_almost_bmi_con4}\textcolor{white}{\eqref{eq_switching_almost_bmi_con4}} \\
    &\textcolor{white}{\mathrm{\st{}~}}\rank(\mathcal{X}_{\xi}) = 1,~\forall \xi  \label{eq_switching_almost_bmi_con5}\textcolor{white}{\eqref{eq_switching_almost_bmi_con5}} \\&\textcolor{white}{\mathrm{\st{}~}}\mathcal{E}_{K,\Ell} = [\mathcal{A}_{K},\mathcal{B}_{K,\Ell}],~\mathcal{F}_{K,\Ell} = \mathcal{B}_{K,\Ell}+\mathbf{Q}_{\mathrm{rob}}^{-1}\Psi_{ck}(\mathcal{X}_{c})^{\top} \\
    &\textcolor{white}{\mathrm{\st{}~}}\mathcal{Y}_{bk} = \mathbf{Q}_{\mathrm{rob}}^{-1}+\Phi_{b}(\mathcal{X}_{b})+\Phi_{k}(\mathcal{X}_{k})+\Pi_{bk}(\mathcal{X}_{bk})
\end{align}
\end{subequations}
where $(\xi,\Xi)$ is given by $(\xi,\Xi) \in \{(b,\mathcal{B}_{K,\Ell}),(c,(\mathcal{C}_{K},\mathcal{K}_{\Ell,K})),\allowbreak(k,\mathcal{X}_c),(bc,(\mathcal{X}_b,\mathcal{X}_c)),(bk,(\mathcal{X}_b,\mathcal{X}_k))\}$ and
\begin{itemize}[leftmargin=*]
   \item $\phi_b$, $\Phi_b$, $\Phi_c$, and $\Phi_k$: Linear functions $(\mathcal{B}_{K,\Ell})_K(\mathcal{B}_{K,\Ell})_K^{\top}$, $\mathcal{B}_{K,\Ell}\mathcal{B}_{K,\Ell}^{\top}$, $\mathcal{C}_K^{\top}\mathcal{C}_K$, and $\mathbf{Q}_{\mathrm{rob}}^{-1}\mathcal{C}_K^{\top}\mathcal{K}_{\Ell,K}^{\top}\mathcal{K}_{\Ell,K}\mathcal{C}_K\mathbf{Q}_{\mathrm{rob}}^{-1}$ expressed in terms of lifting variables $\mathcal{X}_{b}$, $\mathcal{X}_{c}$, and $\mathcal{X}_{k}$, respectively
    \item $\Psi_{bc}$: Linear function $\mathcal{B}_{K,\Ell}^{\top}\Phi_{c}(\mathcal{X}_{c})\mathcal{B}_{K,\Ell}$ expressed in terms of lining variable $\mathcal{X}_{bc}$
    \item $\Psi_{ck}$: Linear function $\mathcal{C}_{K}^{\top}\mathcal{K}_{\Ell,K}^{\top}$ expressed in terms of lining variable $\mathcal{X}_{c}$
    \item $\Pi_{bk}$: Linear function $\mathcal{B}_{K,\Ell}\mathcal{K}_{\Ell,K}\mathcal{C}_K\mathbf{Q}_{\mathrm{rob}}^{-1}\mathcal{C}_K^{\top}\mathcal{K}_{\Ell,K}^{\top}\mathcal{B}_{K,\Ell}^{\top}$ expressed in terms of lifting variable $\mathcal{X}_{bk}$
    \item $\matvec$: vectorization function.
\end{itemize}
\end{theorem}
\begin{proof}
Writing~\eqref{eq_switching_nonlinear} in the epigraph form~\cite[pp. 134]{citeulike:163662}, we get
\begin{align}
    \min_{\mathcal{B}_{K,\Ell},\mathcal{C}_{K},\mathcal{K}_{\Ell,K}}~\gamma
    \mathrm{~\st{}~}\lambda_{\max}(\mathcal{M}) \leq \gamma\mathrm{~and~the~constraints~of~\eqref{eq_switching_nonlinear}}\nonumber
\end{align}
where $\mathcal{M} = \mathcal{E}_{K,L}^{\top}(\mathcal{L}_{\Ell,K}^{-1})^{\top}\mathbf{Q}_{\mathrm{rob}}(\mathcal{L}_{\Ell,K}^{-1})\mathcal{E}_{K,L}$ and $\mathcal{L}_{\Ell,K} = \mathbb{I}-\mathcal{B}_{K,\Ell}\mathcal{K}_{\Ell,K}$. Since $\lambda_{\max}(\mathcal{M}) \leq \gamma \Leftrightarrow \lambda_i(\mathcal{M}-\gamma\mathbb{I}) \leq 0,~\forall i \Leftrightarrow \mathcal{M}-\gamma\mathbb{I} \preceq 0$, where $\lambda_i(\cdot)$ denotes the $i$th eigenvalue, applying Schur's complement lemma~\cite[pp. 650-651]{citeulike:163662} twice to $\mathcal{M}-\gamma\mathbb{I} \preceq 0$ yields
\begin{align}
    \label{eq_schur_proof}
    \left[\begin{smallmatrix}
        \mathcal{Y}_{bk} & \mathcal{E}_{K,\Ell} & \mathcal{B}_{K,\Ell}+\mathbf{Q}_{\mathrm{rob}}^{-1}\mathcal{C}_{K}^{\top}\mathcal{K}_{\Ell,K}^{\top} \\
        \mathcal{E}_{K,\Ell}^{\top} & \gamma\mathbb{I} & \mathbb{O} \\
        \mathcal{B}_{K,\Ell}^{\top}+\mathcal{K}_{\Ell,K}\mathcal{C}_{K}\mathbf{Q}_{\mathrm{rob}}^{-1} &\mathbb{O} & \mathbb{I}
    \end{smallmatrix}\right] \succeq 0~~~~
\end{align}
where $\mathcal{Y}_{bk} = \mathbf{Q}_{\mathrm{rob}}^{-1}+\mathcal{B}_{K,\Ell}\mathcal{B}_{K,\Ell}^{\top}+\mathbf{Q}_{\mathrm{rob}}^{-1}\mathcal{C}_{K}^{\top}\mathcal{K}_{\Ell,K}^{\top}\mathcal{K}_{\Ell,K}\mathcal{C}_{K}\mathbf{Q}_{\mathrm{rob}}^{-1}+\mathcal{B}_{K,\Ell}\mathcal{K}_{\Ell,K}\mathcal{C}_{K}\mathbf{Q}_{\mathrm{rob}}^{-1}\mathcal{C}_{K}^{\top}\mathcal{K}_{\Ell,K}^{\top}\mathcal{B}_{K,\Ell}^{\top}$. 

Even with the relation~\eqref{eq_schur_proof}, the problem is still nonlinear. We thus introduce additional decision variables as follows (\ie{}, liftings, see, \eg{},~\cite{BoydRankRelaxation,rank_constrained_kyotoU} for details) to affinely handle the nonlinear terms without losing the equivalence of the reformulation:
\begin{align}
    &\mathcal{X}_{\xi} = \matvec(\Xi)\matvec(\Xi)^{\top} \Leftrightarrow~\mathrm{\eqref{eq_switching_almost_bmi_con4}~and~\eqref{eq_switching_almost_bmi_con5}}
\end{align}
where $(\xi,\Xi)$ is given by $(\xi,\Xi) \in \{(b,\mathcal{B}_{K,\Ell}),(c,(\mathcal{C}_{K},\mathcal{K}_{\Ell,K})),\allowbreak(k,\mathcal{X}_c),(bc,(\mathcal{X}_b,\mathcal{X}_c)),(bk,(\mathcal{X}_b,\mathcal{X}_k))\}$. Rewriting the constraints of~\eqref{eq_switching_nonlinear} and~\eqref{eq_schur_proof} using these additional lifted variables completes the proof.
\end{proof}
Since the only nonlinearities of the problem~\eqref{eq_switching_almost_bmi} are due to the rank constraints~\eqref{eq_switching_almost_bmi_con5}, we can readily apply a large number of convex rank minimization tools available in the literature, including, but not limited to,~\cite{BoydRankRelaxation,rank_constrained_kyotoU,rank_constrained_boyd,rank_constrained_microsoft,rank_constrained_pourdue_ran}. Also, dropping the rank constraints leads to the following convex optimization with a well-known relaxation approach.
\begin{corollary}
\label{corollary_shorrelax}
The optimization problem in Theorem~\ref{thm_Afixed} without the rank constraints~\eqref{eq_switching_almost_bmi_con5} is the convex relaxation of the problem~\eqref{eq_switching_nonlinear} with Shor's relaxation~\cite{shor_original}, which provide lower bounds for the solutions to the respective problems. 
\end{corollary}
See Corollary~2 of~\cite{stealthy_regret}. 
\begin{proof}
\end{proof}
Since $\mathcal{B}_{K,\Ell}$, $\mathcal{C}_{K}$, and $\mathcal{K}_{\Ell,K}$ are linear in $B_k$, $C_k$, and $\kappa_{\ell,k}$, defined in Remark~\ref{ref_remnote} and in~\eqref{eq_feedback_control}, the optimization problem of Proposition~\ref{corollary_shorrelax} is also convex in these decision variables.
For cases where $A$ also depends on the network topology, the problem~\eqref{eq_switching_almost_bmi} becomes nonlinear due to $(S_{k_1}\cdots S_{k_2})$ terms in $\mathcal{A}_K$ and $\mathcal{B}_{K,\Ell}$. We could still use the lifting approach in the proof of Theorem~\ref{thm_Afixed} if we view $\mathcal{A}_K$ as a decision variable. This problem is still nonlinear in $A_k$

If a feasible solution to the problem~\eqref{eq_switching_nonlinear} is available, we could instead directly solve it using nonlinear optimization approaches, \eg{}, sequential convex programming (SCP)~\cite{scp_trust}, thereby iteratively obtaining a sub-optimal solution with added computational cost.
\subsection{Network Topology Optimization}
We are now ready to propose a robust and optimal network topology switching problem for the ZDA detection, just by expressing the optimization problem of Theorem~\ref{thm_Afixed} in terms of the network topology $\nu$ as in~\eqref{eq_switching_nonlinear}.
\begin{proposition}
\label{proposition_network}
Suppose that Assumptions~\ref{assump_system} and~\ref{assump_affine_topology} hold. The network topology constraints due to the distributed communications and physical connections can be incorporated affinely into \eqref{eq_switching_almost_bmi} of Theorem~\ref{thm_Afixed}, which provide the decision variables of~\eqref{eq_switching_almost_bmi} consistent, with the network topology $\nu$. Furthermore, a Lyapunov-type stability constraint can be added to~\eqref{eq_switching_almost_bmi} while preserving its convex structure.  
\end{proposition}
\begin{proof}
The decision variables of the problems~\eqref{eq_switching_almost_bmi} include $\mathcal{B}_{K,\Ell}$, $\mathcal{C}_{K}$, and $\mathcal{K}_{\Ell,K}$, which are linear in the actuation matrices $B_k$, sensor matrices $C_k$, and control gains $\kappa_{\ell,k}$ of Lemma~\ref{lemma_discretization} and~\eqref{eq_feedback_control}. Therefore, we can affinely set the entries of $B_k$, $C_k$, and $\kappa_{\ell,k}$ with no communications/physical connections to zero if the respective entries of the possible set of all the adjacency matrices $\mathcal{T}_{k}$ in Assumption~\ref{assump_affine_topology} (which can be time-varying, hence the subscript $k$) are zero.

For the Lyapunov-type stability constraint with a given Lyapunov function $V_k = x_{k}^{\top}P_kx_{k}$, we have $x_{k+1}^{\top}P_{k+1}x_{k+1} \leq \alpha x_{k}^{\top}P_kx_{k}$ with $\alpha \in (0,1)$, which gives
\begin{align}
    &\left[\begin{smallmatrix}
    \alpha \mathbb{E}_{k}^{\top}P_k\mathbb{E}_{k} & A_{cl,k}^{\top} \\
    A_{cl,k} & P_{k+1}^{-1}
    \end{smallmatrix}\right] \succeq 0,~k = 0,\cdots,K-1\label{eq_stability} \\
    &A_{cl,k} = \mathbb{E}_{k+1}\mathcal{A}_{K}\mathbb{E}_{0}+\mathbb{E}_{k+1}\mathcal{B}_{K,\Ell}\mathcal{K}_{\Ell,K}\mathcal{C}_{K}
\end{align}
where $\mathbb{E}_{k}$ is a constant matrix defined as $x_k= \mathbb{E}_{k}\mathbf{x}_{K}$ and the controller~\eqref{eq_feedback_control} is used with $\mathbf{v}_{\Ell} = 0$ and $\mathbf{a}_{\Ell} = 0$ in the systems of Lemma~\ref{lemma_discretization}. Using the same lifting variables given in the proof of Theorem~\ref{thm_Afixed}, the stability constraint~\eqref{eq_stability} reduces to a convex constraint. 
\end{proof}

Proposition~\ref{proposition_network} combines Theorem~\ref{thm_Afixed} with network topology and stability constraints while preserving the convexity in~\eqref{eq_switching_almost_bmi}. In particular, this allows for optimal topology switching without affecting the tracking and synchronization performance of the network with distributed communications and physical connections, which will be demonstrated in Sec.~\ref{sec_example}. Again, if $A_k$ depends on the network topology, the problem becomes nonlinear and can be solved by nonlinear optimization tools, including SCP~\cite{scp_trust}.
\section{Numerical Example}
\label{sec_example}
\begin{figure*}
    \centering
    \includegraphics[height=60mm]{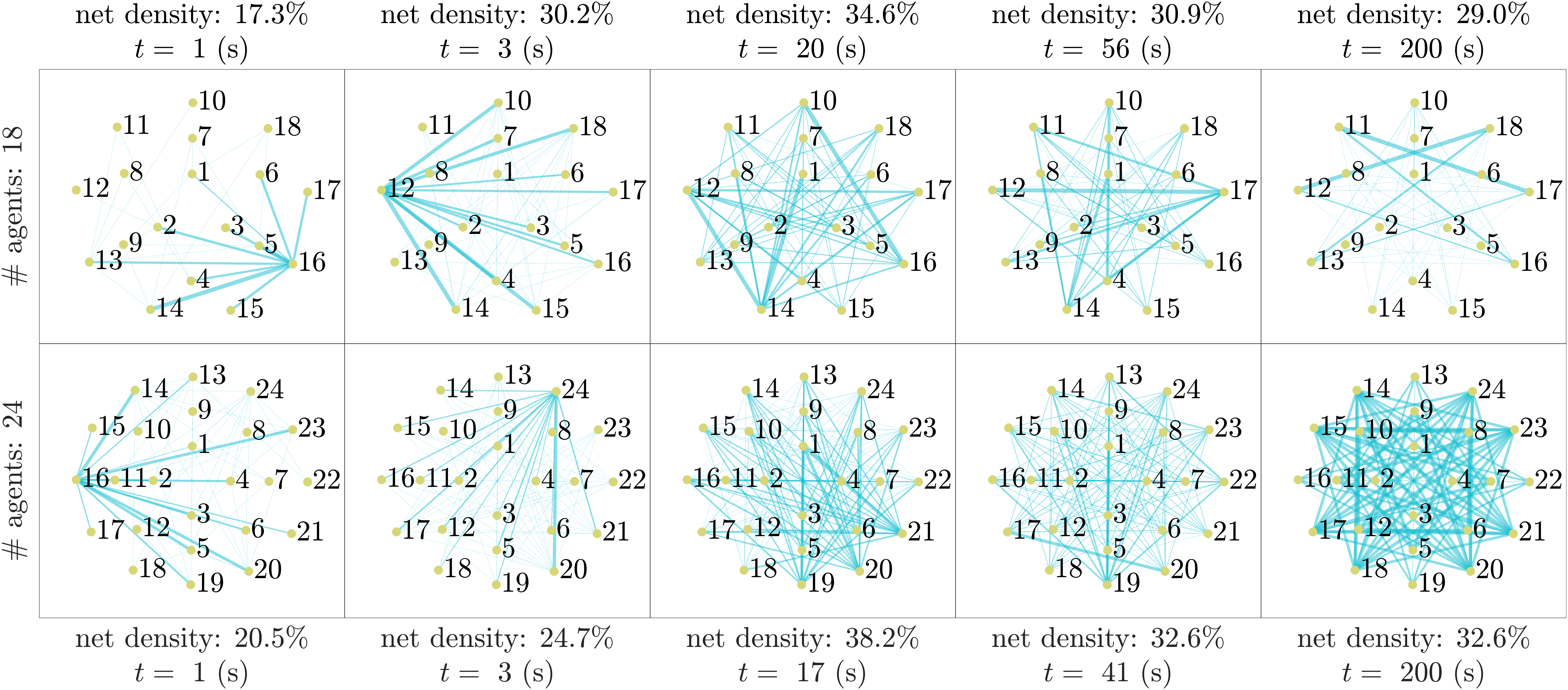}\vspace{-10pt}
    \caption{$5$ representative optimal topologies out of $23$ switchings for $18$ agents (first row) and $5$ representative optimal topologies out of $30$ switchings for $24$ agents (second row), where line weights represent weights of communication edges.}
    \label{switching_fig}\vspace{-10pt}
\end{figure*}
This section demonstrates the robust and optimal network topology switching of Sec.~\ref{sec_example} for networked double integrators. As seen in Example~\ref{ex_dinet}, this network is subject to ZDAs even when the system is both controllable and observable.

\subsubsection*{Setup}
We consider the networked double integrators with $18$ agents and $24$ agents. The switching problem of Proposition~\ref{proposition_network} is solved with the rank minimization approach of~\cite{rank_constrained_kyotoU} every time the feasible set of the adjacency matrices $\mathcal{T}_k$ changes, where the time horizon is selected as $K = 2$. The set $\mathcal{T}_k$ in Proposition~\ref{proposition_network} is determined by the physical positions of the mobile agents with given maximum communication radii at each time step $k$. The network density (\ie{}, the number of the edges divided by the maximum possible number of the edges) is constrained to be less than $40$\% while ensuring the graph is undirected and strongly connected. This example illustrates the network's communication switching. The reconfiguration of the physical connections is not considered here.

The agents can exchange their relative state measurements if there is an edge between them and only the leader agent has access to its absolute state measurement. The hierarchical approach of~\cite{Ref:phasesync} is used for the feedforward controller $\mathbf{v}_{\Ell}$ of~\eqref{eq_feedback_control} and its contraction theory-based tracking and synchronization constraints are also used for the stability constraint of Proposition~\ref{proposition_network}. The asynchronous control actuation and sensing sampling periods $\Delta t_u$ and $\Delta t_y$ are selected as $\Delta t_u = 5.00\times 10^{-1}$~(sec) and $\Delta t_y = 1.00$~(sec). For attack detection, we simply use the residual filter of~\cite{attack_bullo}, assuming that we have a countermeasure that rejects attacks once they are detected. The Gaussian process noise and sensor noise with the standard deviation $1.00\times 10^{-4}$ and $5.00\times 10^{-3}$, respectively, are added to each state and measurement of each agent.

\subsubsection*{Discussion}
\begin{figure}
    \centering
    \includegraphics[height=62.88mm]{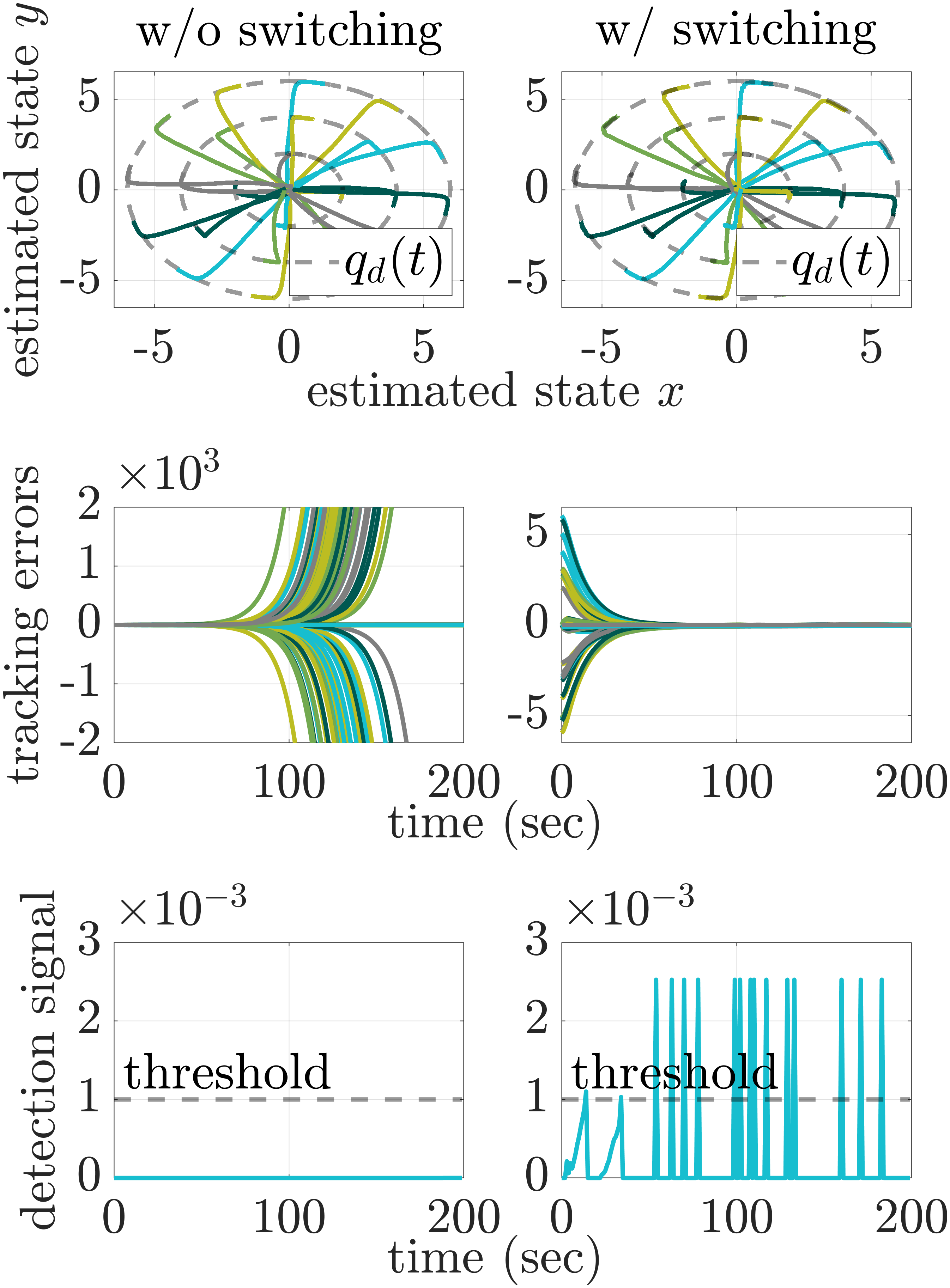} 
    \vspace{-10pt}
    \caption{Position trajectories (first row), tracking errors of attacked states (second row), and detection signals of residual filters (third row) without topology switching (first column) and with topology switching (second column), where $x$ and $y$ are horizontal and vertical positions of double integrators and $q_d(t)$ are given target trajectories.}
    \label{comparison_fig}
\end{figure}

The simulation setup results in $23$ switchings for $18$ agents and $30$ switchings for $24$ agents, and each row of Figure~\ref{switching_fig} illustrates the $5$ representative optimal topologies among these, respectively. As implied in Table~\ref{tab_results}, these switchings indeed reduce the attack robustness metric while guaranteeing that the controllability, observability, and minimum attack sensitivity metrics are non-zero. This fact can be visually appreciated also in Fig.~\ref{comparison_fig}, which shows the simulation results for the networked double integrators attacked by ZDAs at random time steps with probability. ZDAs are constructed as described in Sec.~\ref{sec_zda_definitions}.
\begin{enumerate}
\item The first row of Fig.~\ref{comparison_fig} indicates that the system is controllable and observable for both cases with and without switchings due to non-zero $J_{\mathrm{con}}$ and $J_{\mathrm{obs}}$.
\item The second row indicates that the attacked states for the network equipped with the optimal switchings converge to their desired trajectories even with the process noise, sensor noise, and ZDAs due to non-zero $J_{\mathrm{sen}}$ and small $J_{\mathrm{rob}}$. The tracking errors diverge without switchings. 
\item The third row indicates that the ZDAs at random time steps can be detected for the network with the optimal switchings due to non-zero $J_{\mathrm{sen}}$. The ZDAs remain stealthy for the case without switchings.
\end{enumerate}
\begin{table}[htbp]\vspace{-5pt}
\caption{Performance metrics of Sec~\ref{sec_metrics} averaged over time.\label{tab_results}}
\begin{center}
\begin{tabular}{ c c c c c }
\hline
\hline
 & $J_{\mathrm{con}}$ & $J_{\mathrm{obs}}$ & $J_{\mathrm{rob}}$ & $J_{\mathrm{sen}}$ \\
\hline
$18$ agents & $2.47\times 10^{-2}$ & $1.54\times 10^{-4}$ & $4.05$ & $1.10\times 10^{-6}$ \\
$24$ agents & $2.47\times 10^{-2}$ & $1.40\times 10^{-4}$ & $4.06$ & $1.09\times 10^{-6}$ \\ \hline
\makecell{w/o switching \\($24$ agents)} & $2.47\times 10^{-2}$ & $7.56\times 10^{-2}$ & $6.89$ & $1.80\times 10^{-14}$ \\
\hline
\hline
\end{tabular}
\end{center}\vspace{-20pt}
\end{table}
\section{Conclusion}
This paper proposes a novel framework for robust and optimal topology switching in networked linear control systems, specifically addressing intrinsic, sampling, and enforced ZDAs. We utilize a performance metric explicitly expressed in terms of the network communication topology, which aims to minimize the classical $\mathcal{H}_{\infty}$ norm while capturing controllability, observability, and attack sensitivity via constraints. The key contribution of our method is the transformation of the metric minimization problem into an equivalent rank-constrained optimization problem, offering an efficient convex optimization-based approach to implementing robust optimal topology switching. Our simulations of networked linear systems show that optimal topology switching can robustly and optimally detect and mitigate ZDAs while maintaining system controllability and observability. Future work would include extensions of our framework to nonlinear systems with uncertainty, consideration of sensor attacks, and incorporation of learning and data-driven methods for evolving stealth attack patterns.

\bibliographystyle{IEEEtran}
\bibliography{main}

\end{document}